\documentstyle[12pt]{article}

\begin{document} 
\begin{flushright}
OITS 765.1\\
July 2005
\end{flushright}
\vspace*{-3cm}
\begin{flushleft}{To be submitted to the \\
Brief Report section of the\\
Phys.\ Rev.\ C.}
\end{flushleft}
\vspace*{1cm}

\begin{center}  {\Large {\bf Associated Particle Distributions in
Jets Produced in Heavy-Ion Collisions}}
\vskip .75cm
 {\bf   Rudolph C. Hwa$^1$ and  Zhiguang Tan$^{1,2}$}
\vskip.5cm

 {$^1$Institute of Theoretical Science and Department of
Physics\\ University of Oregon, Eugene, OR 97403-5203, USA\\
\bigskip
$^2$Institute of Particle Physics, Hua-Zhong Normal University,
Wuhan 430079, P.\ R.\ China}
\end{center}
\vskip.5cm

\begin{abstract} 
We study the pion distributions in jets produced in heavy-ion collisions in association with trigger particles restricted to definite momentum ranges. We determine the centrality dependence of the associated particle distributions for both d+Au and Au+Au collisions in the recombination model. The central-to-peripheral ratios show significant dependence on centrality for Au+Au collisions, but  negligible dependence for d+Au collisions. The results are in agreement with the qualitative features of the data.
\vskip0.5cm
PACS numbers:  25.75.Dw,  25.75.Gz
\end{abstract}

In an earlier paper \cite{hy1} dihadron correlation in jets produced in heavy-ion collisions (HIC) was studied in the recombination model \cite{hy2}. In that paper the distribution of particles ($\pi^+$ at $p_2$) associated with a trigger ($\pi^+$ at $p_1$) was calculated; it is the conditional distribution for every fixed $p_1$, denoted by
\begin{eqnarray}
\left.{ dN_{\pi^+} \over  dp_2}\right|_{\pi^+(p_1)} = { dN_{\pi^+\pi^+} /dp_1 dp_2   \over dN_{\pi^+}/ dp_1  } \ .
\label{1}
\end{eqnarray}
In \cite{hy1} this equation is expressed in Eq.\ (18) there in a form that exposes the hard parton integration. Equation (\ref{1}) was then integrated over $p_1$ for trigger momentum in the range $4<p_1<6$ GeV/c, corresponding to one of the trigger windows of the STAR experiment \cite{st} at the Relativistic Heavy-Ion Collider (RHIC). However, that integrated result does not correspond to the quantity that is presented by STAR as the measured associated particle distribution (APD). Instead of the integral of the ratio in Eq.\ (\ref{1}), the experimental APD thus far analyzed is the ratio of the integrals
\begin{eqnarray}
{dN_{\rm assoc}\over p_2dp_2}={\int_4^6dp_1 p_1 \rho_2(1,2) \over \int_4^6dp_1 p_1 \rho_1(1)},   \label{2}
\end{eqnarray}
where 
\begin{eqnarray}
\rho_1(1)={dN\over p_1dp_1},  \hskip 2cm \rho_2(1,2)={d^2N\over p_1p_2dp_1dp_2}.   \label {3}
\end{eqnarray}
The ratio in Eq.\ (\ref{2}) was calculated in \cite{ht} for central Au+Au collisions, and was shown to be in reasonable agreement with the data when due consideration of the differences in the particle types measured is taken into account. In this short paper we show the centrality dependence of the APD in Eq.\ (\ref{2}) for both d+Au and Au+Au collisions, as well as for two different ranges of the trigger momentum.

It should be noted that if the ranges of integration over $p_1$ in Eq.\ (\ref{2}) are taken to be infinitesimal, there is no essential difference between Eqs.\ (\ref{1}) and (\ref{2}). There is nothing wrong with Eq.\ (\ref{1}), which corresponds to the usual conditional probability often considered in statistical physics. The APD in Eq.\ (\ref{2}) with a significant  range of 2 GeV/c for the trigger $p_T$ is preferred for data analysis mainly to gain better experimental statistics. Over the intermediate $p_T$ region where the calculated APDs decrease by three orders of magnitudes, they do not differ by very much in the general shape whether  Eqs.\ (\ref{1}) or (\ref{2}) is used. However, the ratio of the distributions for central to peripheral collisions depends more sensitively on the trigger range. The calculation of that ratio according to Eq.\ (\ref{2}) is the main objective of this work.

As in \cite{hy1}, we consider here only the correlation between two pions, $\pi^+\pi^+$, one being the trigger particle in two trigger windows, the other being the associated particle in the range $p_2<p_{\rm trigger}$. This is a simpler and  trackable system than what has been measured, which includes all charged hadrons, but is rich enough in the various recombination components to provide a meaningful representation of the APD. In our calculation we use the basic formulas in the recombination model \cite{hy1,hy2}, for which $\rho_1(1)$ is
\begin{eqnarray}
{dN_{\pi_1}\over p_1dp_1}={\xi\over p}\sum_{ij}\int dk k f_i(k)\left[{1\over p^2}\int dq_1{\cal T}(p_1-q_1)S_i^j\left({q_1\over k}\right)+{1\over k} D_i^{\pi_1}\left({p\over k}\right)\right],  \label{4}
\end{eqnarray}
where the thermal-thermal recombination component ($\cal TT$) has been omitted, since we study here the particles in a jet.  In Eq.\ (\ref{4}) $f_i(k)$ is the distribution of hard partons in HIC and is proportional to the number of binary collisions $\left<N_{\rm coll}\right>$; $\xi$ is the average fraction of hard partons that emerge from the dense medium to hadronize outside. ${\cal T}(q)$ is the thermal parton distribution \cite{hy1,hy2}, whose parametrization for Au+Au collisions has now been extended to all centralities \cite{ht}. $S_i^j$ is the shower parton distribution (SPD) given in \cite{hy3}. $D_i^\pi$ is the fragmentation function from hard parton $i$ to pion and accounts for the shower-shower recombination component. For $\rho_2(1,2)$ in Eq.\ (\ref{2}) we have 
\begin{eqnarray}
{d^2N_{\pi_1\pi_2}\over p_1dp_1p_2dp_2}={1\over p_1^2p_2^2}\int\left(\prod_{i=1}^4 {dq_i\over q_i}\right)F_4(q_1,q_2,q_3,q_4)R_{\pi_1}(q_1,q_3,p_1)R_{\pi_2}(q_2,q_4,p_2),  \label{5}
\end{eqnarray}
where $R_\pi$ is the recombination function for the formation of a pion, and $F_4$ is 4-parton distribution, which in our case here is 
\begin{eqnarray}
F_4=({\cal TS+SS})_{13}({\cal TS+SS})_{24},   \label{6}
\end{eqnarray}
where the $\cal TT$ terms have been omitted for the same reason as in Eq.\ (\ref{4}). The details of how $\cal S$ and $(\cal SS)$ are expressed in terms of SPDs are described in \cite{hy1, hy2, ht}.

For d+Au collisions the hard parton distribution $f_i(k)$ and the parameters of the thermal distributions at various centralities are given in \cite{hy4}, while for Au+Au collisions the corresponding quantities are given in \cite{ht}. We now can calculate the APD in Eq.\ (\ref{2}) for d+Au collisions, as well as for Au+Au collisions, for all centralities. The results for $dN/dp_2$  are shown in Figs.\ 1 and 2 for trigger windows $4<p_1<6$ GeV/c and $6<p_1<8$ GeV/c, respectively. What is notable in these figures is that there is negligible  centrality dependence in (a) for d+Au collisions, but significant dependence in (b) for Au+Au collisions. The ratios dAu(cent)/dAu(peri) and AuAu(cent)/AuAu(peri) are shown in Fig.\ 3 as functions of the momentum of the associated particle. The solid lines in Fig.\ 3 are for Au+Au collisions and are far from 1, indicating appreciable dependence on centrality. At $p_T\approx 1$ GeV/c the ratio is about 3, which implies that the thermal partons in central collisions drastically enhance the production of hadrons.  The data shown in triangles are from STAR \cite{ja} for AuAu(0-5\%)/pp for trigger in $4<p_T^{trig}<6$ GeV/c with all charged hadrons included in the associated particles. Our result agree with the data for $p_T<3$ GeV/c. Other analysis of the STAR data \cite{yg}  are also in support of our result. The large factor of 3 around $p_T=1$ GeV/c would be hard to attain by the medium modification of the fragmentation function, if fragmentation is considered instead of recombination for hadronization.

The dashed lines for d+Au collisions are just above 1, and are virtually indistinguishable from one another in the two trigger window cases.  The data shown are preliminary results from PHENIX on the ratio of the APDs for 0-20\% d+Au to p+p collisions for all charged hadrons on the near side in association with a pion trigger in the  $p_T$ range of 5-10 GeV/c \cite{ng}. Despite the differences in the calculated and measured quantities, the general agreement between the two suggests that our model calculation has captured the essence of the nature of the correlation. There are also other preliminary experimental data on conditional yield in d+Au collisions that show approximate independence on centrality  \cite{jb}. 

The reason why there is less centrality dependence in d+Au collisions is that the thermal-shower component plays a much less prominant role than it does in Au+Au collisions. Since the parts in the factors $\xi f_i(k)$ in $\rho_1(1)$ and $\rho_2(1,2)$ in Eq.\ (\ref{2}) that depend on $\left<N_{\rm coll}\right>$ cancel each other, only the $\cal T$ terms in Eqs.\ (\ref{4}) and (\ref{6}) produce the centrality dependence, and they are minor in d+Au collisions. For Au+Au collisions $\cal SS$ recombination is insignificant in the intermdediate $p_T$ region, so the $\cal TS$ terms are dominant in Eqs.\ (\ref{4}) and (\ref{6}), resulting in the dependence on centrality shown in part (b) of  Figs.\ 1 and 2.  Figure 3 exhibits the difference of the two colliding systems in terms of ratios. 

We have calculated the associated particle distributions only for $p_2>1$ GeV/c because our formalism for particle production is not reliable for $p_T<1$ GeV/c. For that reason we do not integrate the associated particle distribution to determine the yield per trigger, since it is dominated by the contribution from the low $p_T$ region. Nevertheless, by inspection of the distributions in Figs.\ 1 and 2, it is clear that the yield/trigger is not likely to depend on centrality very much for d+Au collisions, but significantly for Au+Au collisions.

Our conclusion is that when the associated particle distributions are calculated as the ratio of the integrals given in Eq.\ (\ref{2}), we find significant dependence on centrality for Au+Au collisions \cite{ja,yg}, but negligible dependence on centrality for d+Au collisions, and is in agreement with the data \cite{ng,jb}.

 This work was supported, in part,  by the U.\ S.\ Department of Energy under
Grant No. DE-FG03-96ER40972.  

\newpage
\begin{center}
\section*{Figure Captions}
\end{center}

\begin{description}
\item
Fig.\ 1. Associated particle distribution for $\pi^+$ as trigger in the range $4<p_T<6$ GeV/c and for $\pi^+$ as the associated particle in the range $1<p_T<5$ GeV/c for all centralities in (a) d+Au collisions and (b) Au+Au collisions.

\item
Fig.\ 2. Associated particle distribution for $\pi^+$ as trigger in the range $6<p_T<8$ GeV/c and for $\pi^+$ as the associated particle in the range $1<p_T<7$ GeV/c for all centralities in (a) d+Au collisions and (b) Au+Au collisions.
 
 \item
Fig.\ 3. Ratios of AuAu(central)/AuAu(peripheral) and dAu(central)/dAu(peripheral) as functions of the momentum of associated $\pi^+$ for two trigger windows.
 The data in triangles are from \cite{ja} for AuAu(0-5\%)/pp with trigger in $4<p_T^{\rm trig}<6$ GeV/c. The shaded region shows the systematic uncertainties.   The data in squares are preliminary results from PHENIX \cite{ng} for dAu(0-20\%)/pp for all charged hadrons on the near side in association with pion trigger in the interval $5<p_T<10$ GeV/c.

\end{description}

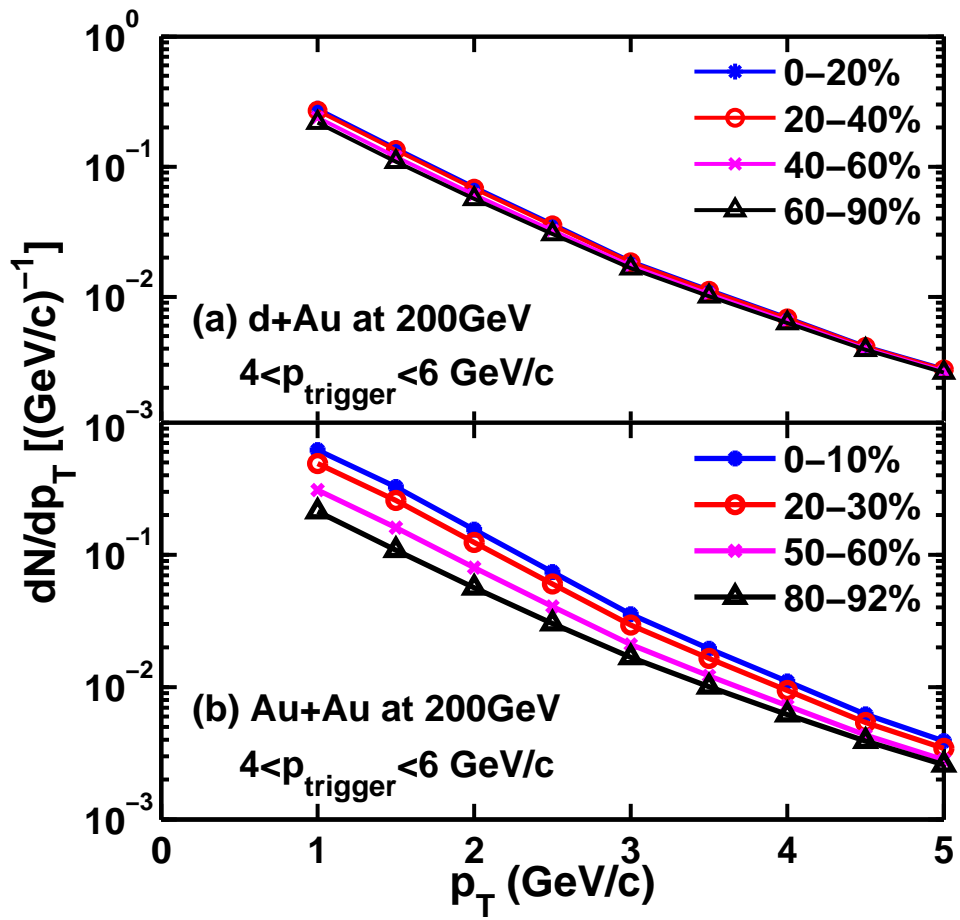
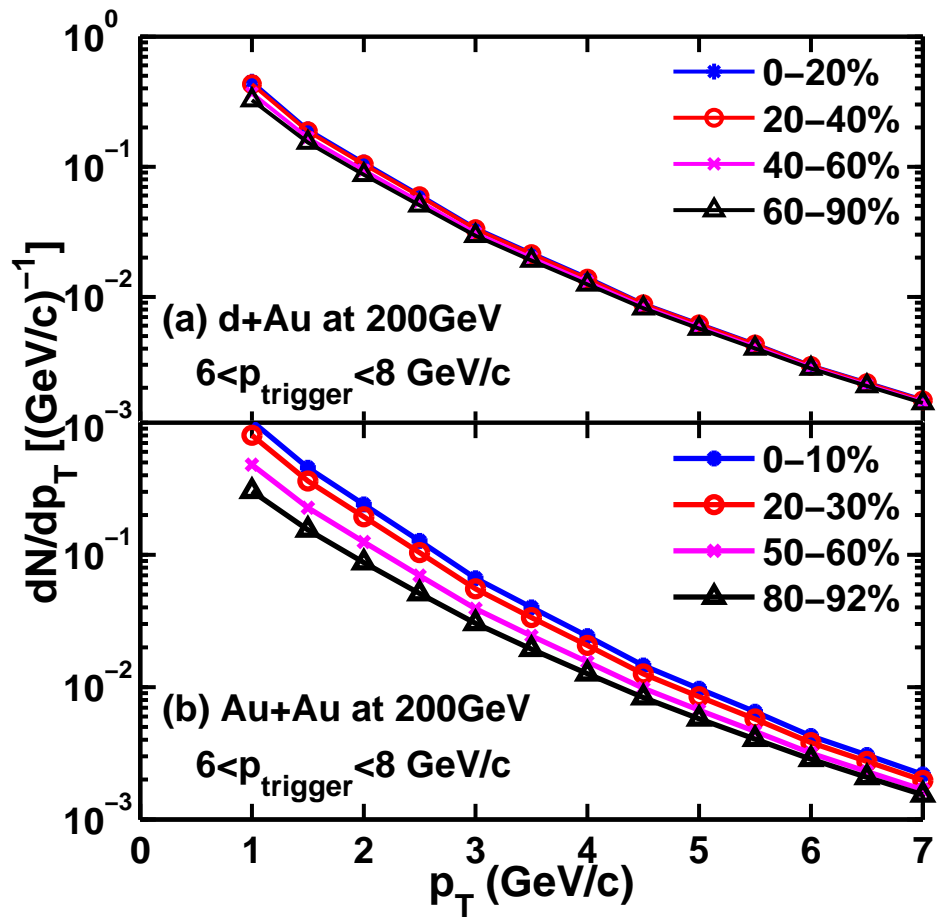
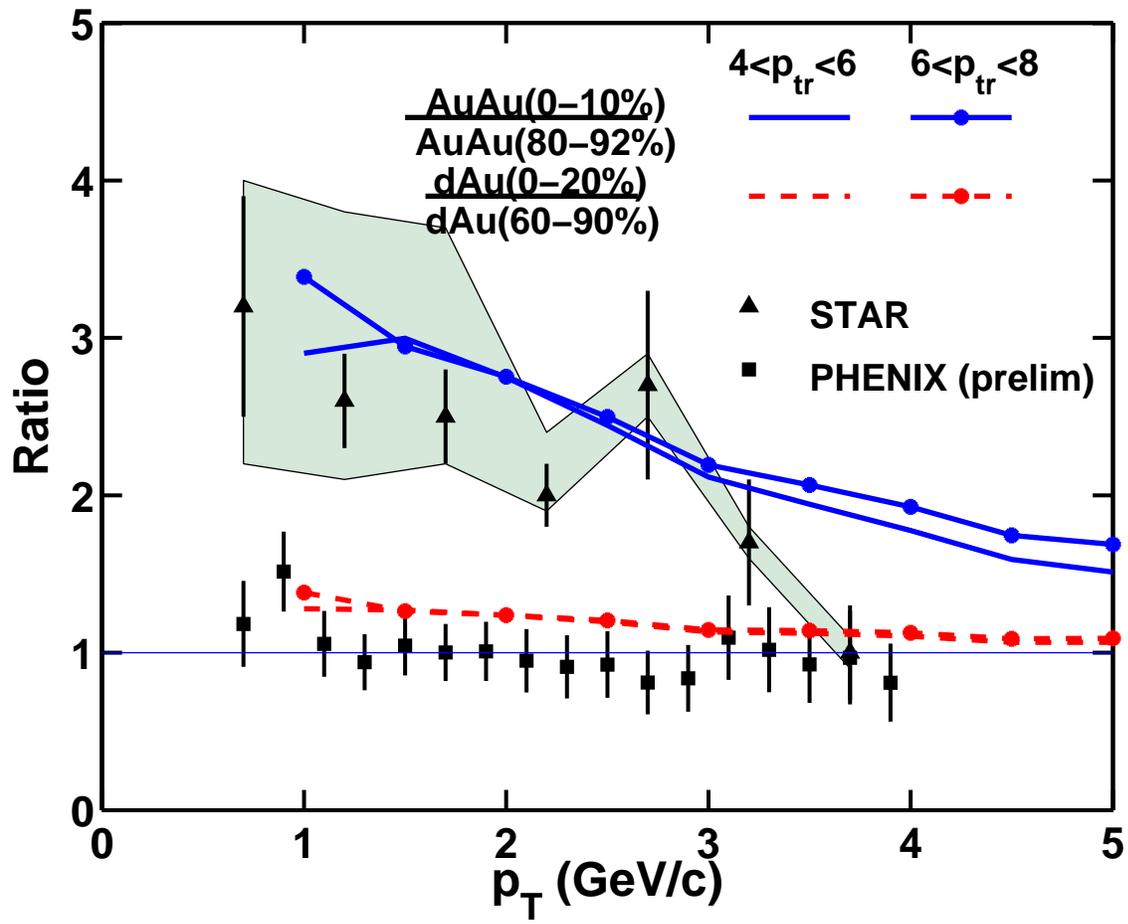
\end{document}